# GPU-based ultra-fast direct aperture optimization for online adaptive radiation therapy

**Chunhua Men, Xun Jia, and Steve B. Jiang**

Department of Radiation Oncology, University of California San Diego, La Jolla, CA 92037, USA

E-mail: sbjiang@ucsd.edu

Online adaptive radiation therapy (ART) has great promise to significantly reduce normal tissue toxicity and/or improve tumor control through real-time treatment adaptations based on the current patient anatomy. However, the major technical obstacle for clinical realization of online ART, namely the inability to achieve real-time efficiency in treatment re-planning, has yet to be solved. To overcome this challenge, this paper presents our work on the implementation of an intensity modulated radiation therapy (IMRT) direct aperture optimization (DAO) algorithm on graphics processing unit (GPU) based on our previous work on CPU. We formulate the DAO problem as a large-scale convex programming problem, and use an exact method called column generation approach to deal with its extremely large dimensionality on GPU. Five 9-field prostate and five 5-field head-and-neck IMRT clinical cases with 5×5 mm$^2$ beamlet size and 2.5×2.5×2.5 mm$^3$ voxel size were used to evaluate our algorithm on GPU. It takes only 0.7~2.5 seconds for our implementation to generate optimal treatment plans using 50 MLC apertures on an NVIDIA Tesla C1060 GPU card. Our work has therefore solved a major problem in developing ultra-fast (re-)planning technologies for online ART.

## 1. Introduction

Radiation therapy is traditionally a static process, whereby treatment plans are generated based on a snapshot of the patient's anatomy prior to treatment, and then delivered over a number of weeks. Patient's anatomy, however, as a dynamic system, can vary significantly from fraction to fraction. This inter-fraction anatomical variation can severely compromise the success of radiation therapy (Antolak *et al.*, 1998; Lee *et al.*, 2004; Nichol *et al.*, 2007; Chan *et al.*, 2008; van der Wielen *et al.*, 2008). Image guided radiation therapy (IGRT), offering online volumetric imaging via CT on-rails (Kokubo *et al.*, 2002; Wong *et al.*, 2001) or on-board cone beam CT (CBCT) (Jaffray and Siewerdsen, 2000; Jaffray *et al.*, 2002), provides a platform for the development of online adaptive radiation therapy (ART) that allows real-time treatment adaptations based on the current patient anatomy (Wu *et al.*, 2002; Mohan *et al.*, 2005; Wu *et al.*, 2008; Godley *et al.*, 2009). The adjustment of patient position through IGRT before each treatment fraction cannot solve the problem posed by tumor and/or organ regression, deformation and their relative position change, where a change to the patient's treatment plan is needed. A number of studies have evaluated the dosimetric impact of mid-treatment re-planning, indicating that a significant percentage of patients could have the target coverage improved and have the normal tissue toxicity reduced (Yan *et al.*, 1997; Ghilezan *et al.*, 2004; van de Bunt *et al.*, 2006; Kuo *et al.*, 2006; Langen *et al.*, 2006; Yan, 2008).

To generate an intensity modulated radiation therapy (IMRT) plan based on daily CBCT scan while the patient is lying in the treatment position, three key components have to be done in real time: 1) anatomy segmentation, 2) dose calculation, and 3) treatment plan (re-)optimization. To accelerate the computation process, one way is to use computer clusters or traditional supercomputers. However, they are expensive and not readily available to most clinical users. Computer graphics processing unit (GPU), on the other hand, armed with hundreds of processing cores, can be effectively used for parallel computing, and is affordable to clinical users. GPU offers a potentially powerful computational platform for convenient and affordable high-performance computing in a clinical environment (Li *et al.*, 2007; Samant *et al.*, 2008; Xing *et al.*, 2008; Hissoiny *et al.*, 2009; Jacques *et al.*, 2009; Preis *et al.*, 2009; Wallner, 2009; Jia *et al.*, 2010). For the anatomy segmentation, we have implemented and evaluated of various demons deformable image registration algorithms on GPU and 256×256×100 pulmonary 4D CT images require 7 seconds for segmentation (Gu *et al.*, 2010); For the dose calculation, we have developed GPU-based ultra-fast dose calculation using a finite size pencil beam model and the computational time for calculating dose deposition coefficients for a 9-field prostate IMRT plan with this new framework is less than 1 second (Gu *et al.*, 2009); For the treatment plan (re-)optimization, we have developed an ultra fast fluence map optimization (FMO) algorithm on GPU and for a 9-field prostate IMRT plan of 5×5 mm$^2$ beamlet size and 2.5×2.5×2.5 mm$^3$ voxel size, and we can finish the re-optimization in 2.8 seconds (Men *et al.*, 2009).

In this paper, we will focus on IMRT treatment plan (re-)optimization for online ART using GPU-based direct aperture (re-)optimization (DAO) method. Traditionally, IMRT treatment plans are developed using a two-stage process. FMO problem must be followed by a leaf-sequencing stage in which the fluence maps are decomposed into a manageable number of apertures that are deliverable using a multi-leaf collimator (MLC) system. A major drawback in the decoupling of the treatment planning problem into an FMO problem and an MLC leaf-sequencing problem is that there is a potential loss in the treatment quality. To overcome this, researchers have developed approaches that integrate the beamlet based FMO and leaf-sequencing problems into a single optimization model, which are usually referred to as direct aperture optimization (DAO) approaches. Meanwhile, quality assurance (QA) of MLC leaf sequences is a challenge in online ART. DAO is able to directly optimize MLC apertures that can be easily delivered thus the necessity for MLC QA is greatly reduced.

Even though DAO problems are well studied, most of them focus on heuristic search methods, such as simulated annealing (Shepard *et al.*, 2002; Earl *et al.*, 2003; Zhang *et al.*, 2006; Mestrovic *et al.*, 2007) and genetic method (Li *et al.*, 2003). These search algorithms in general are inefficient. Another method has been developed to solve DAO problem in a deterministic way (Romeijn *et al.*, 2005; Men *et al.*, 2007) which used a column generation method to handle its large dimensionality and the optimal treatment plan can be obtained in about 2 minutes. To further improve its efficiency for online treatment re-planning applications, we have implemented this method on GPU. This paper describes the implementation of column generation method on GPU to solve IMRT DAO problem. It also presents the evaluation of the ability of our approach to efficiently generate the high-quality treatment plans for online ART.

**2. Methods and Materials**

*2.1 Direct Aperture Optimization Model*

We will denote the set of deliverable apertures by $K$. In IMRT optimization, each beam is decomposed into a set of beamlets (denoted by $N$) and a particular setting of beamlets $A_k \in N$ forms aperture $k \in K$. With each aperture $k$ we associate a decision variable $y_k$ that indicates the intensity of that aperture. The set of voxels that represents the patient's CT image is denoted by $V$. In addition, we denote the dose to voxel $j \in V$ by $z_j$. The voxel dose $z_j$ ($j \in V$) is calculated using a linear function of the intensities of the apertures through the so-called dose deposition coefficients $D_{kj}$, the dose received by voxel $j \in V$ from aperture $k \in K$ at unit intensity

$$z_j = \sum_{k \in K} D_{kj} y_k. \qquad (1)$$

$D_{kj}$ can be obtained from the following equation:

$$D_{kj} = \sum_{i \in A_k} D_{ij} \qquad (2)$$

where $D_{ij}$ represents the dose received by voxel $j \in V$ from beamlet $i \in N$ at unit intensity. We can obtain $D_{ij}$'s for a 9-field prostate IMRT clinical case using our in-house dose calculation engine implemented on GPU (Gu *et al.*, 2009) in 1 second.

Our DAO model employs treatment plan evaluation criteria that are quadratic one-sided voxel-based penalties. If we denote the set of target voxels by $V_T$, we can write the criteria as:

$$F_j^-(z_j) = \alpha_j \left(\max\{0, T_j - z_j\}\right)^2 \quad j \in V_T \tag{3}$$

$$F_j^+(z_j) = \beta_j \left(\max\{0, z_j - T_j\}\right)^2 \quad j \in V. \tag{4}$$

Here $\alpha_j$ and $\beta_j$ represent the penalty weights for underdosing and overdosing penalty, respectively and $T_j$ represents the penalty threshold for voxel $j$. If we let

$$F(\mathbf{z}) = \sum_{j \in V}[F_j^-(z_j) + F_j^+(z_j)] \tag{5}$$

our model can be written as:

$$\min F(\mathbf{z})$$

Subject to

$$z_j = \sum_{k \in K} D_{kj} y_k \quad j \in V \tag{6}$$

$$y_k \geq 0 \quad k \in K.$$

If $x_i$ denotes the intensity of beamlet $i \in N$ and $K_i$ denotes set of apertures in which the beamlet $i$ is included, the following relationship between the sets of decision variables has to hold

$$x_i = \sum_{k \in K_i} y_k \quad i \in N. \tag{7}$$

*2.2 The optimization algorithm*

The model is convex and the decision variables are the intensities of all MLC deliverable apertures. Mathematically, the DAO model is the same as the FMO model developed in previous work (Fox *et al.*, 2008; Men *et al.*, 2009) in which the decision variables are the intensities of all beamlets. While we may use the same gradient projection method as in our previous work to solve DAO problem, it is neither practical nor necessary. It is clear that the number of deliverable apertures is huge. Since the FMO model is already a challenging large-scale optimization problem (number of decision variables is around $10^3 \sim 10^4$), the DAO model is intractable due to the huge dataset (number of decision variables is more than $10^{18}$). On the other hand, we only need to select 30~80 apertures for deliverability consideration, even if we were able to solve such a huge problem. We therefore use a column generation method (see, *e.g.*, (Bazaraa *et al.*, 2006)) which iteratively employs a sub-problem whose optimal solution either provides a suitable aperture to add to a given pool of allowable apertures or concludes that the current solution is optimal (Romeijn *et al.*, 2005; Men *et al.*, 2007). The flowchart for implementing our DAO algorithm is shown in Figure 1. The master problem is to find the optimal intensities of this given pool of apertures.

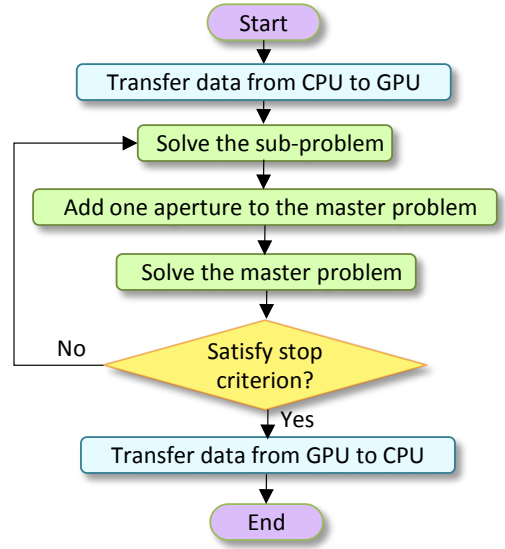

**Figure 1.** Flowchart of GPU-based column generation method for solving the DAO problem.

*2.3 GPU implementation*

Recent advancement of general purpose GPU technologies offers incredible resources for parallel computing with affordable price. The programming approach we used was compute unified device architecture (CUDA) which was developed by NVIDIA and performs scientific calculations on GPU as a data-parallel computing device.

To solve the DAO model using column generation method, we need to solve the master problem and the sub-problem iteratively (see Figure 1). In each iteration, an aperture candidate is added to the pool of apertures. The master problem is to find the optimal intensities for these selected apertures. Note the objective function is convex quadratic in our master problem, thus the direction of steepest descent is that of the negative gradient. However, since the decision variables, which are the intensities of selected apertures, should be nonnegative, moving along the steepest descent direction may lead to infeasible intensities. We therefore use the gradient projection method (see, *e.g.*, (Bazaraa *et al.*, 2006)) which projects the negative gradient in a way that improves the objective function while maintaining feasibility. We have implemented gradient projection algorithm on GPU to solve the FMO problem for IMRT treatment (re-)planning (Men *et al.*, 2009) and we will use it to solve our master problem in this work.

In each iteration, the sub-problem identifies one aperture which decreases the objective value most if added to the master problem. We will denote the set of beams by $B$ and the set of apertures that can be delivered by a MLC system from beam direction $b \in B$ by $K_b$. Let us denote the dual multiplier associated with constraints $z_j = \sum_{k \in K} D_{kj} y_k$ by $\pi_j$ for $j \in V$. For the concept of dual multiplier, please refer to text books related to optimization techniques (see, *e.g.*, (Bazaraa *et al.*, 2006)). The sub-problem is to solve the following problem,

$$\min_{k \in K_b} \sum_{i \in A_k} \left( \sum_{j \in V} D_{ij} \pi_j \right). \tag{8}$$

Before solving this sub-problem, we can compute the value placed in parentheses, $w_i \equiv \sum_{j \in V} D_{ij} \pi_j$, for each beamlet $i \in N$ explicitly. Assume that there are $n$ beamlets in each row of an MLC, our problem then becomes that, for each row of MLC finding a consecutive set of beamlets for which the sum of their $w_i$ values is minimized. In fact, we can find such a set of beamlets for a given row by searching through the $n$ beamlets from left to right only once. At an intermediate step of this searching process, let the cumulative value of $w_i$ over all beamlets considered so far, the maximum cumulative value found so far, and the best value found so far be denoted by $v$, $\bar{v}$ and $v^*$, respectively. In addition, let $l$ and $r$ denote the left and right MLC leaf positions at current searching step and $l^*$ and $r^*$ denote the best left and right MLC leaf positions so far, respectively. We can then solve this problem through a polynomial-time algorithm on GPU, whose correctness has been proven (Bates and Constable, 1985; Bentley, 1986). The algorithm can be summarized as follows:

*Initialization*
   $v = \bar{v} = v^* = 0; l = l^* = 0; r = r^* = 1.$
*Main Iterative Loop*
   1. $v = v + w_r$.
   2. If $v > \bar{v}$, then $\bar{v} = v$ and $l = r$.
   3. If $v - \bar{v} < v^*$, then $v^* = v - \bar{v}$, $l^* = l$ and $r^* = r + 1$.
   4. If $r < n$, then $r = r + 1$, go to Step 1; else stop.

In our CUDA implementation, each thread independently runs the algorithm for an MLC row in parallel to obtain $l^*$, $r^*$ and $v^*$ for that row. Then summation of $v^*$ for all MLC rows within each beam can be calculated. The aperture from the beam which attains the smallest summation values will be the solution to the sub-problem (8), and it is then added to the pool of apertures.

The dose deposition coefficient matrix is a sparse matrix due to the fact that a specified aperture only contributes to limited number of voxels. We therefore store the $D_{kj}$'s in a *compressed sparse row* (CSR) format, which is the most popular general-purpose sparse matrix representation. CSR stores column indices and nonzero values in arrays indices and data. An additional array of row pointers is also necessary in the CSR representation. In each iteration of our column generation method as shown in Figure 1, the dose deposition coefficient sparse matrix and the matrix transpose have to be updated (because a new aperture is added to the model). While transposing a sparse matrix might be easy for the traditional series CPU computing, it is quite a challenge for parallel GPU computing. To improve the performance of our CUDA code, we used a newly developed template library *Thrust* from NVIDIA (Hoberock and Bell, 2009) to re-organize (sorting, counting, *etc.*) the data and then obtain the matrix transpose.

## 3. Experiments and Results

To test our implementation, we used five clinical cases of prostate cancer (Case P1~P5) and five clinical cases of head-and-neck cancer (Case H1~H5). For prostate cancer cases, nine 6 MV co-planar beams were evenly distributed around the patient and the prescription dose to planning target volume (PTV) was 73.8 Gy. For head-and-neck cases, five 6 MV co-planar beams were evenly distributed around the patient and the prescription dose to PTV1 was 73.8 Gy, and the prescription dose to PTV2 was 54 Gy. PTV1 consists the gross tumor volume (GTV) expanded to account for both sub-clinical disease as well as daily setup errors and internal organ motion; PTV2 is a larger target that also contains high-risk nodal regions and is again expanded for same reasons. For all cases, we used a beamlet size of 5×5 mm$^2$ and voxel size of 2.5×2.5×2.5 mm$^3$ for target and organs at risk (OARs). For unspecified tissue (*i.e.*, tissues outside the target and OARs), we increased the voxel size in each dimension by a factor of 2 to reduce the optimization problem size. The full resolution was used when evaluating the treatment quality (does volume histograms (DVHs), dose color wash, isodose curves, *etc*.). Table 1 shows the dimensions of these 10 cases in the DAO models.

| Case | # beamlets | # voxels | # non-zero $D_{ij}$'s | Running time (sec) |
|------|-----------|----------|----------------------|--------------------|
| P1   | 7,196     | 45,912   | 2,763,243            | 1.7                |
| P2   | 7,137     | 48,642   | 2,280,076            | 0.7                |
| P3   | 5,796     | 28,931   | 1,765,294            | 0.8                |
| P4   | 7,422     | 39,822   | 2,717,424            | 2.3                |
| P5   | 8,640     | 49,210   | 3,086,884            | 1.6                |
| H1   | 5,816     | 33,252   | 1,576,418            | 1.0                |
| H2   | 8,645     | 59,615   | 3,162,752            | 2.4                |
| H3   | 9,034     | 74,438   | 3,500,188            | 1.8                |
| H4   | 6,292     | 31,563   | 1,596,168            | 1.8                |
| H5   | 5,952     | 42,330   | 2,215,202            | 2.5                |

**Table 1.** Case dimensions and GPU running time on an NVIDIA Tesla C1060 GPU card for direct aperture plan optimization implementations.

We tested our CUDA implementation on an NVIDIA Tesla C1060 GPU card, which has 30 multiprocessors (each with 8 SIMD processing cores) and 4GB of memory. In order to have a fair comparison, we generated 50 apertures for all test cases. The running time for DAO plan optimization for these cases on a GPU is shown in Table 1. The amount of time required ranges in 0.7~2.5 seconds. It takes 2~3 minutes to solve such problems on an Intel Xeon 2.27 GHz CPU.

We then analyzed the accuracy of the results. The obtained results on GPU are very similar to those obtained on CPU with only $10^{-2} \sim 10^{-3}$ relative difference in objective values. The difference can be attributed to single floating point precision on GPU and is negligible in clinical practice. In fact, by checking the DVHs, dose color wash, and isodose curves, no differences could be observed between CPU and GPU results. Figure 2 and 3 show the DVHs and isodose/dose color wash superimposed on representative CT slices of a prostate cancer case (Case P1) and of a head-and-neck cancer case (Case H1), corresponding to optimal DAO treatment plans, respectively.

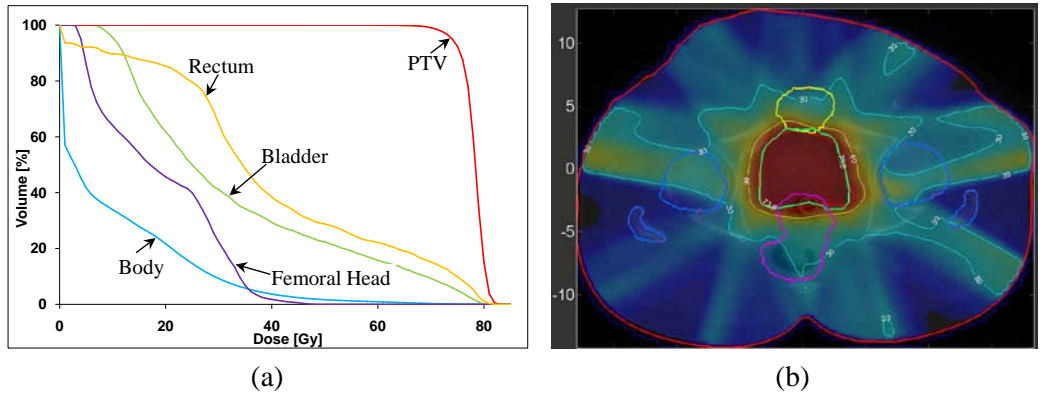

**Figure 2**. The optimal treatment plan obtained from our GPU-based DAO implementation for Case P1 (prostate cancer): (a) DVHs; (b) dose color wash /isodose on a representative CT slice.

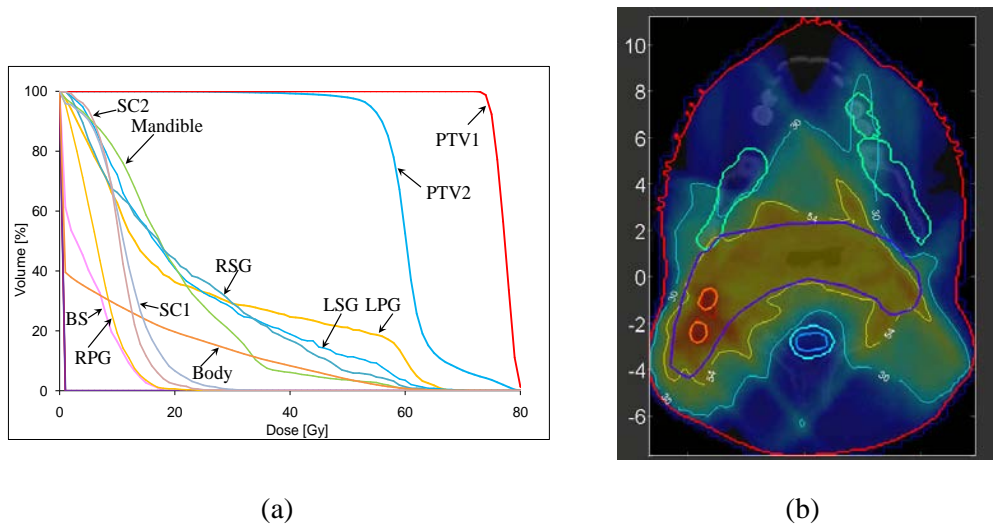

**Figure 3**. The optimal treatment plan obtained from our GPU-based DAO implementation for Case H1 (head-and-neck cancer): (a) DVHs; (b) dose color wash /isodose on a representative CT slice. (BS: brain stem; SC1: spinal cord 1; SC2: spinal cord 2; LSG: left submandibular gland; RSG: right submandibular gland; LPG: left parotid gland; RPG: right parotid gland)

## 4. Discussion and Conclusions

In this paper, we presented the implementation of a GPU-based ultra-fast DAO algorithm in IMRT treatment (re-)planning. Instead of using heuristic search algorithms, we used a direct aperture optimization approach to design radiation therapy treatment plans for individual patients. We tested our implementation on five 9-field prostate IMRT cases and five 5-field head-and-neck IMRT cases. Our results showed that an optimal plan can be obtained in 0.7~2.5 seconds using a Tesla C1060 card which is suitable for on-line ART.

We also investigated the performance of our CUDA implementation on different GPUs, including NVIDIA's GeForce 9500 GT and GTX 285. The GeForce 9500 GT has only 4 multiprocessors and produced limited speedup (1-3x). The GTX 285 has 30 multiprocessors (same as C1060) and delivered similar speedup results as shown in Table 1. We found, for these test cases, that the speedup factors solely depend on the number of multiprocessors per GPU.

Our model can be used for both initial treatment plan optimization on planning CT images and for treatment plan re-optimization on daily CBCT images. In Equation (3) and (4), $\alpha_j$ and $\beta_j$ represent the penalty weights for underdosing and overdosing penalty, respectively, and $T_j$ represents the penalty threshold for voxel $j$. If we denote the set of targets by $T$ and the set of critical structures by $C$, the set of structures by $S = T \cup C$, and the set of voxels in structure $s \in S$ by $v_s$. Then Equation (3) and (4) can be reformulated as follows:

$$F_{s-}(z) = \frac{\alpha_s}{|v_s|} \sum_{j \in v_s} \left( \max\{0, T_s^- - z_j\} \right)^2 \qquad s \in T \qquad (9)$$

$$F_{s+}(z) = \frac{\beta_s}{|v_s|} \sum_{j \in v_s} \left( \max\{0, z_j - T_s^+\} \right)^2 \qquad s \in S \qquad (10)$$

where $\alpha_s$ and $\beta_s$ are penalty weights, and $T_s^-$ and $T_s^+$ are the penalty thresholds for structure $s$. Criteria (9) penalize underdosing below the underdosing threshold $T_s^-$ in all targets $s \in T$ while Criteria (10) penalize overdosing above the overdosing threshold $T_s^+$ in all structures $s \in S$. In this case, the model is to optimize the original treatment plan. On the other hand, in Equation (3) and (4), if we let $\alpha_j = \beta_j = 1$ and $T_j$ (for all $j \in V$) represents the dose distribution from the original treatment plan, deformed from the planning CT to the daily CBCT, then in this case, we re-optimize the treatment plan by trying to reproduce the dose distribution from the original plan within a tolerance to accommodate the changed patient's geometry. The discussion of this re-optimization mode can be found in our previous work (Men *et al.*, 2009).

We also evaluated the efficiency of our algorithm using various beamlet and voxel sizes. We noticed that the running time increased/decreased if the number of voxels increased/decreased by changing voxel resolution. However, It is interesting to see that the GPU running time did not change too much for both prostate and head-and-neck cancer cases even though more/less beamlets were used. For the FMO model, the GPU running time highly depends on the size of the clinical case, *i.e.*, number of beamlets and voxels. For the DAO model, the number of beamlets does not play a critical role. Remember that using column generation method, we iteratively solve a master problem and a sub-problem. Sub-problem is to identify an aperture which decreases the objective function most if added to the master problem. We identify this aperture by passing through beamlets once for each beamlet row and hence the number of beamlets in each row and the number of rows decide the size of the problem. Since we exploited fine-grained parallel algorithm, we were able to solve the sub-problem very fast (<0.1 ms) for all cases and then the effect of solving sub-problem can be negligible to the total GPU running time. In fact, we noticed that solving the master problem is the most time-consuming. For the master problem, the decision variables are the intensities for all selected apertures. All cases generated the same number of apertures, then the efficiency of master problem highly depends on the number of voxels. Therefore, even though more/less beamlets were used in clinical cases, it did not result in longer/shorter GPU running time. Higher beamlet resolution will lead to more accurate treatment plan. Using traditional CPU-based computational tools and/or FMO models, higher beamlet resolution results in longer computational time and this limits the size of beamlets. Using our GPU-based DAO treatment plan (re-)optimization algorithm, we can handle very large number of beamlets without losing efficiency.

We generated 50 apertures for each scenario but it may not be the best number of apertures according to our previous work (Men *et al.*, 2007) in which we have developed 2 stopping rules which decide the necessary number of apertures. The aim of this work is to implement column generation method on GPU instead of investigating this algorithm. In order to have a fair comparison of GPU running time, we used fixed number of apertures. Moreover, we are able to handle various MLC hardware constraints in our optimization model on GPU, including standard consecutiveness, interdigitation, connectedness and jaws-only delivery though we only showed the results for the consecutiveness constraint.

**Acknowledgements**

We would like to thank NVIDIA for providing GPU cards. This work is supported in part by the University of California Lab Fees Research Program.

12					C Men *et al.*Jaffray D A, Siewerdsen J H, Wong J W and Martinez A A 2002 Flat-panel cone-beam computed tomography for image-guided radiation therapy *International Journal of Radiation Oncology Biology Physics* **53** 1337-49

Jia X, Lou Y, Li R, Song W Y and Jiang S B 2010 Cone beam CT reconstruction from undersampled and noisy projection data via total variation *Medical Physics* **37** 1757-60

Kokubo M, Yamashita M, Okumachi H and Hiraoka M 2002 The accuracy of irradiated position in a linac combined with a rail-on CT with a single common couch *International Journal of Cancer* 460-1

Kuo Y C, Wu T H, Chung T S, Huang K W, Chao K S C, Su W C and Chiou J F 2006 Effect of regression of enlarged neck lymph nodes on radiation doses received by parotid glands during intensity-modulated radiotherapy for head and neck cancer *American Journal of Clinical Oncology-Cancer Clinical Trials* **29** 600-5

Langen K M, Meeks S L and Kupelian P 2006 *Integrating new Technologies in the Clinic: Monte Carlo and Image-Guided Radiation Therapy,* ed B H Curran*, et al.* (Madison, WI: Medical Physics Publishing) p 736

Lee C M, Shrieve D C and Gaffney D K 2004 Rapid involution and mobility of carcinoma of the cervix *International Journal of Radiation Oncology Biology Physics* **58** 625-30

Li M, Yang H, Koizumi K and Kudo H 2007 Fast cone-beam CT reconstruction using CUDA architecture *Medical Imaging Technology* 243-50

Li Y J, Yao J and Yao D Z 2003 Genetic algorithm based deliverable segments optimization for static intensity-modulated radiotherapy *Physics in Medicine and Biology* **48** 3353-74

Men C, Gu X, Choi D J, Majumdar A, Zheng Z, Mueller K and Jiang S B 2009 GPU-based ultrafast IMRT plan optimization *Physics in Medicine and Biology* **54** 6565-73

Men C, Romeijn H E, Tas Z C and Dempsey J F 2007 An exact approach to direct aperture optimization in IMRT treatment planning *Physics in Medicine and Biology* **52** 7333-52

Mestrovic A, Milette M P, Nichol A, Clark B G and Otto K 2007 Direct aperture optimization for online adaptive radiation therapy *Medical Physics* **34** 1631-46

Mohan R, Zhang X D, Wang H, Kang Y X, Wang X C, Liu H, Ang K, Kuban D and Dong L 2005 Use of deformed intensity distributions for on-line modification of image-guided IMRT to account for interfractional anatomic changes *International Journal of Radiation Oncology Biology Physics* **61** 1258-66

Nichol A M, Brock K K, Lockwood G A, Moseley D J, Rosewall T, Warde P R, Catton C N and Jaffray D A 2007 A magnetic resonance imaging study of prostate deformation relative to implanted gold fiducial markers *International Journal of Radiation Oncology Biology Physics* **67** 48-56

Preis T, Virnau P, Paul W and Schneider J J 2009 GPU accelerated Monte Carlo simulation of the 2D and 3D Ising model *Journal of Computational Physics* **228** 4468-77
12